\documentclass[11pt,twoside]{article}
\usepackage{graphicx}

\usepackage{asp2006}
\usepackage{lscape}

\def\gtorder{\mathrel{\raise.3ex\hbox{$>$}\mkern-14mu
    \lower0.6ex\hbox{$\sim$}}}
\def\ltorder{\mathrel{\raise.3ex\hbox{$<$}\mkern-14mu
    \lower0.6ex\hbox{$\sim$}}}

\begin{document}
\title{Dark Matter Substructure, Filaments and Assembling Disks}    
\author{Isaac Shlosman}   
\affil{JILA, University of Colorado, Boulder, CO 80309, USA}     

\begin{abstract}  
We review some general properties of assembling galactic dark matter (DM) halos which 
have a direct effect on the baryon dynamics. Specifically, we focus on the mutual
dynamical feedback between baryons and DM which influence disk formation and evolution,
by comparing models with and without baryons evolved from identical initial
conditions. Baryons are found to be capable of modifying the DM density profiles
of inner halos, leading to isothermal profiles rather than the NFW cusps. Those can be 
leveled off by the subsequent dynamical friction of the subhalos at lower $z$. 
Furthermore,
subhalos appear efficient in triggering galactic bars and ablating cold gas from
the disks, and therefore quenching of star formation there.
\end{abstract}

\section{Adding Baryons to Forming Galaxies}    

Formation of luminous parts of galaxies is inherently connected with baryon dynamics in 
the background of the dark matter (DM). While only one sixth of the matter is baryonic
in the WMAP3 Universe (e.g., Spergel et al. 2007), it apparently dominates the inner
regions of galaxies because of the ability to dissipate energy. The associated efficiency
in getting rid of the angular momentum leads to de facto separation between 
baryons and DM. The relevant galactic dynamics probably transcends the simple adiabatic 
contraction (e.g., Blumenthal et al. 1986) and the actual evolution appears to be more complex.
To more fully understand the role of baryons in galaxy evolution, we focus on two issues:
(1) how do baryons affect the background DM in galaxies, and (2) how does the distribution 
of DM on scales larger and smaller than galactic halos influences the accumulation and
dynamics of baryons in the central regions, where disks form and grow. 
In the following, we make a broad use of representative models 
of pure DM (PDM models) and baryon$+$DM (BDM models), which have been evolved from
identical initial conditions (Romano-Diaz et al. 2009a).
The general properties of DM halos which have been established so far lead to 
important consequences for baryon evolution in the galactic centers. Those in turn exert
feedback on the DM, which affects shapes, density profiles, angular momentum and 
substructure. We limit out discussion to those properties which have a direct impact on
disk evolution.

\begin{figure}[t!!!!!]
\begin{center}
\resizebox{6.5cm}{!}{\includegraphics{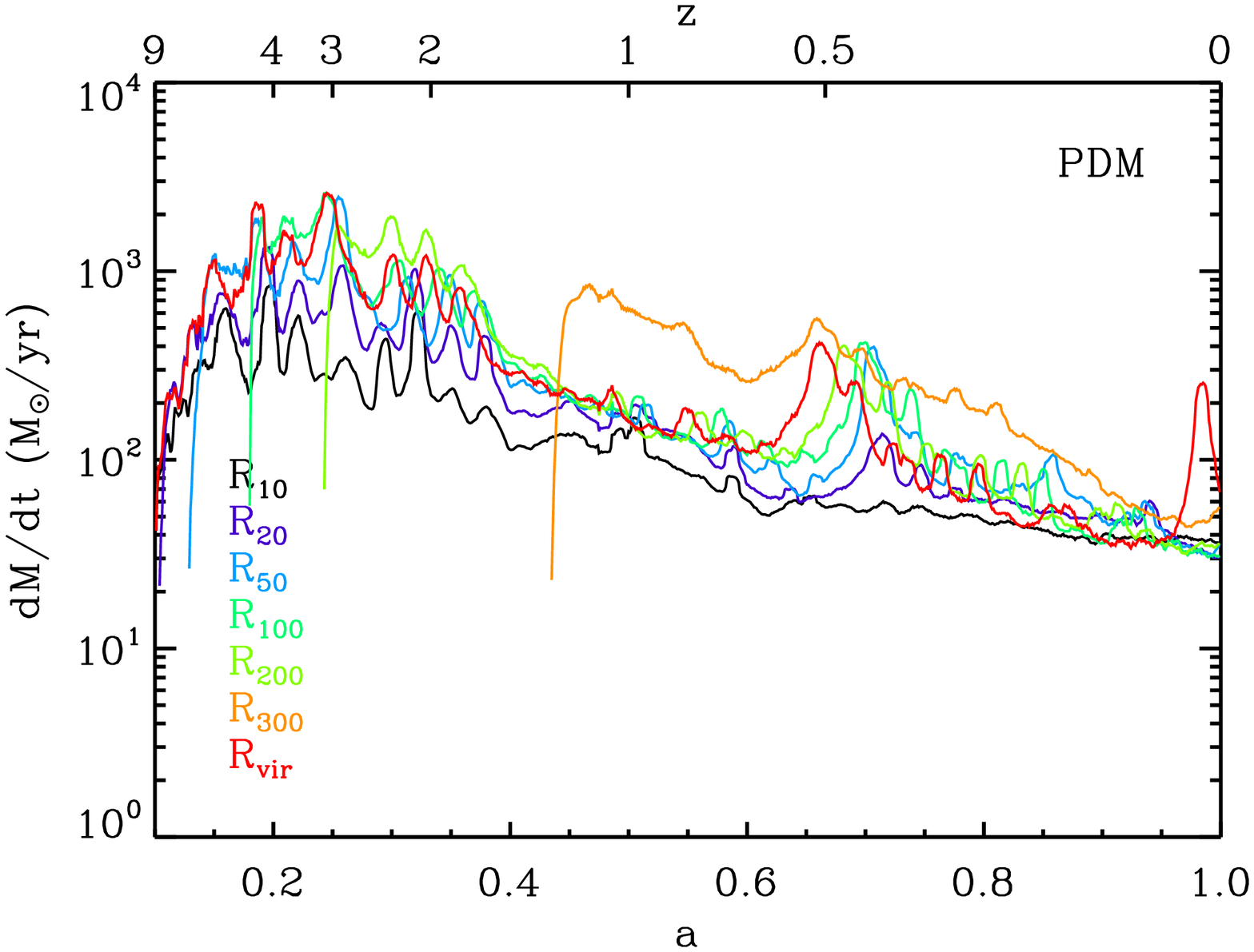}}
\resizebox{6.5cm}{!}{\includegraphics{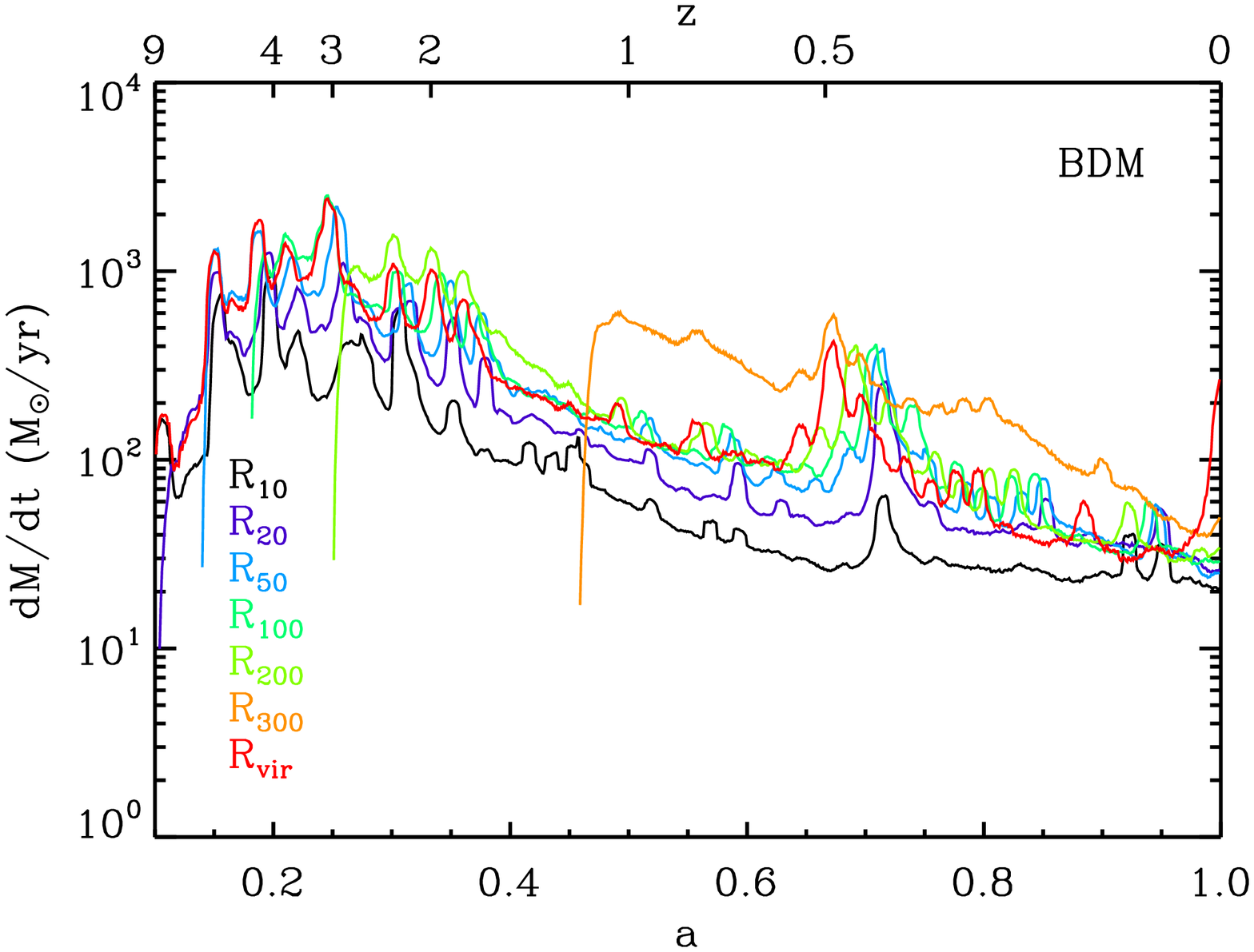}}
\end{center}
\caption{
\footnotesize
DM influx rates along the filaments in the pure DM (left) and baryons$+$DM
(right) models, with identical initial conditions. The color-coded curves correspond 
to the fixed radii of 10~kpc -- 300~kpc and the DM virial radii, as a function of $z$ 
and the cosmological expansion parameter $a$  (Romano-Diaz et al. 2009b).
}
\end{figure}
\begin{figure}[t!!!!!]
\begin{center}
\resizebox{6.7cm}{!}{\includegraphics{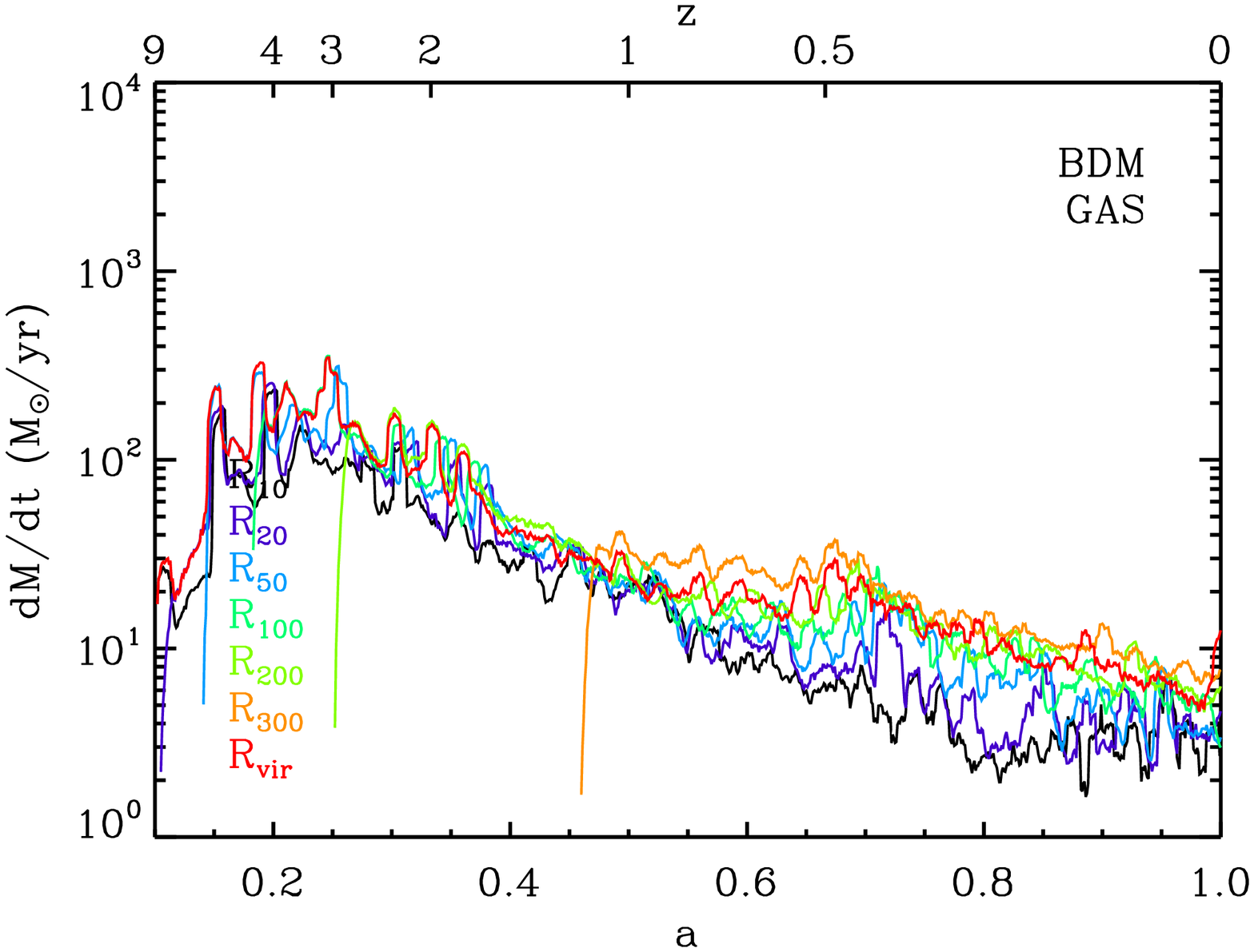}}
\resizebox{6.cm}{!}{\includegraphics{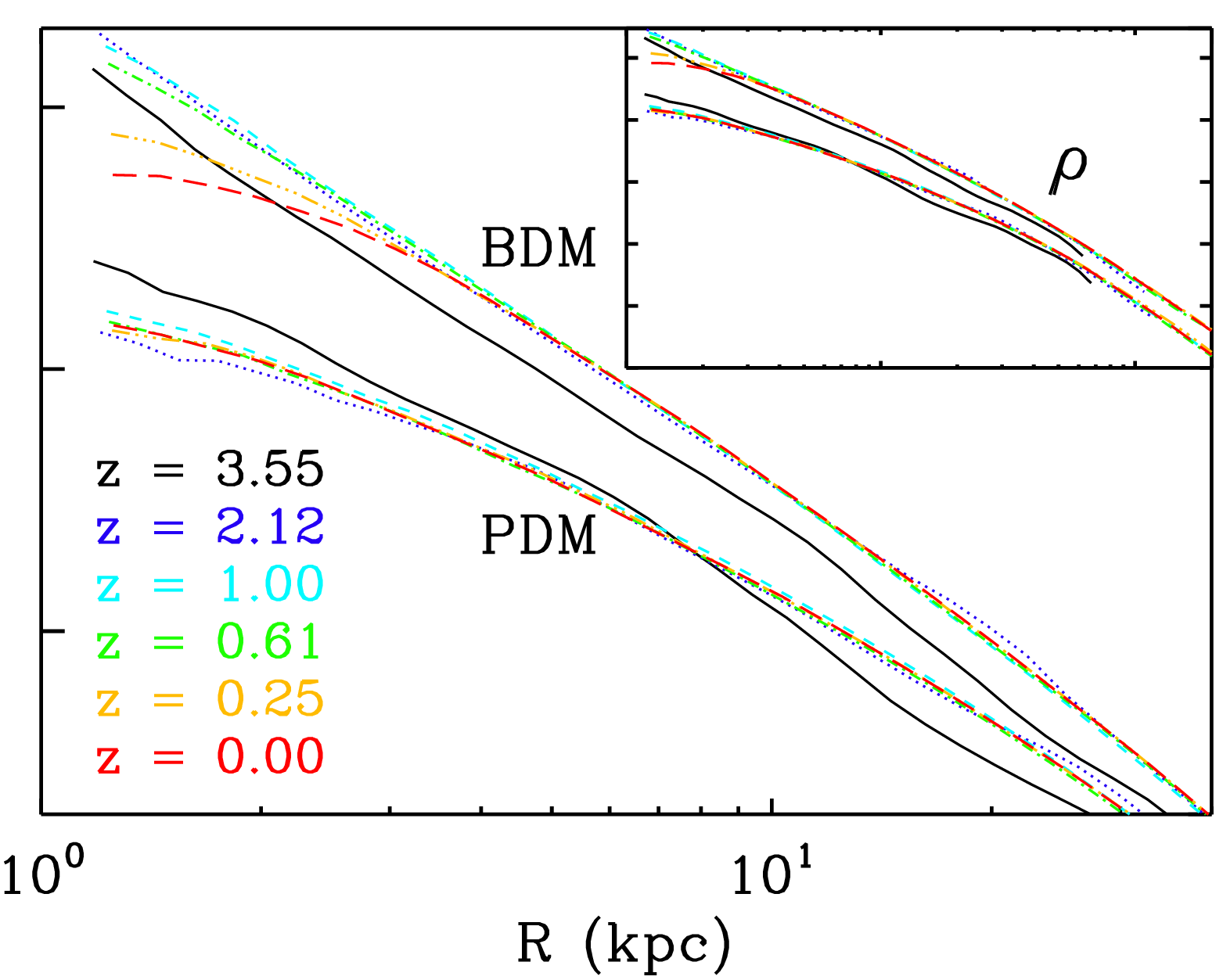}}
\end{center}
\caption{
\footnotesize
\underline{\it Left:} As in Fig.~1 but for the gas inflow along the filaments
(Romano-Diaz et al. 2009b). \underline{\it Right:} Evolution of DM density
profiles $\rho(R)$ within the inner 20~kpc in the PDM and BDM models. The insert
shows $\rho$ within 200~kpc. The curves are displaced vertically for clarity  
(Romano-Diaz et al. 2008a).
}
\end{figure}

Filaments and accretion through filaments emerge as the dominant mode of galaxy growth
(e.g., Dekel et al. 2008). It is important that the cosmological filaments which 
dominate the large-scale DM structure initially extend to the very small scales of a 
few kpc where the seed galactic disks form at intermediate $z$ (Shlosman 2009). The 
characteristic ``cat's cradle'' displays how the inflowing baryons smoothly join the disk, 
experiencing little `shocking' (Fig.~3 in Heller et al. 2007). At small radii, the
clumpy gas influx appears tangent to the protodisk, in apparent repeling action of
the angular momentum. The inflow along
the DM filaments prevents the gas from being virialized before it joints the
disk. Furthermore, it assures that the gas will lose some fraction of its angular momentum 
to a filament. The typical DM accretion rates along the filaments decay with time and
show a mild difference between the PDM and BDM models (Fig.~1a,b).
This difference amounts to a somewhat supressed influx rate in the inner 10~kpc---20~kpc
after $z\sim 1$ and is related to the action of baryons as we discuss below. The
typical gas accretion rate along the filaments fits comfortably to the overall
picture of galaxy growth. Fig.~2a exhibits inflow which is sufficient for a massive
disk formation in the long run and can support the currently observed star formation 
rates in a Milky Way-type galaxies.

\section{DM and Baryon Settling in Pure DM and DM$+$Baryons Models}

The DM halo buildup, in the absence of baryons, is known to lead to a universal density
profile with a characteristic scale $R_s$ corresponding to the logarithmic
slope of --2 (e.g., Navarro et al. 1997, NFW). Within this radius, the density is
cuspy, its slope tends to --1 when $R\rightarrow 0$. The reason for this universality 
is currently debated
but it is important that baryons can substantially modify this density profile
(El-Zant et al. 2001; 2004; Tonini et al. 2006),
bridging the gap with observations. Those hint at a central flat density core rather
than a cusp (e.g., de Blok 2007; Kuzio de Naray et al. 2008). 
Romano-Diaz et al. (2008b) find that the NFW cusp formation 
is by-passed in favor of an isothermal DM cusp, in BDM models (Fig~2b). 
The isothermal cusp in turn can be 
leveled off by the action of penetrating minor mergers after $z\sim 1$. To avoid
the ambiguity in the definition of $R_s$ in the BDM models, we invoke the radius of
a maximal circular velocity, $R_{\rm vmax}$, in the halo instead. The fraction
$\gamma\equiv R_s/R_{\rm vmax}\sim 0.46$ is universal for the NFW fit. For all models
we define ${\tilde R}_s\equiv\gamma R_{\rm vmax}$.   

While formation of isothermal cusps is related to the adiabatic contraction and
involves a certain degree of dissipation by baryons, their demise can be associated
with clumpy accretion of DM subhalos glued by baryons. It happens well beyond
the epoch of major mergers, after $z\sim 1$. Romano-Diaz et al. (2008a,b) have
shown that the late stages of inflow along the filaments involve subhalos
which cluster there, and enter the prime halo before they merge among themselves.
Due to higher binding energy, the subhalos in the BDM models can penetrate
the inner halo. Within the virial radius of the prime halo, the mass function
of subhalos evolves strongly, especially when baryons are present (Romano-Diaz et 
al., in preparation). The mass ratio 
of the most massive subhalo to the prime halo
decreases with time, but is $\sim 8\times 10^{-4}$ even at $z=0$, higher than
the minimum required by El-Zant et al. (2001) for the dynamical friction to
be efficient over orbital times. The influx of subhalos is well
correlated with the `smooth' DM streaming out of the center and `cooling' down. 

Cold, rotationally supported baryons are expected to accumulate within 
${\tilde R}_s$, 
subject to expulsion by stellar or AGN feedback. Some baryons will be trapped in 
the filaments and in shells outside the virial
radius, $R_{\rm vir}$. These shells form when incoming subhalos are tidally 
disrupted but avoid
capture by the gravitational well. Hence, it is not clear at present to what extent
the baryonic content of a galaxy corresponds to the universal fraction
of baryons. For a MW-type galaxy, Romano-Diaz et al. (2009a) find that major mergers 
tend to increase the baryon fraction, while minor mergers decrease it --- the net 
effect appears to be 
negligible. This result is based on a fine-tuning of the stellar feedback parameter,
based on a large number of disk models (Heller et al. 2007).

\begin{figure}[t!!!!!]
\begin{center}
\resizebox{4.65cm}{!}{\includegraphics{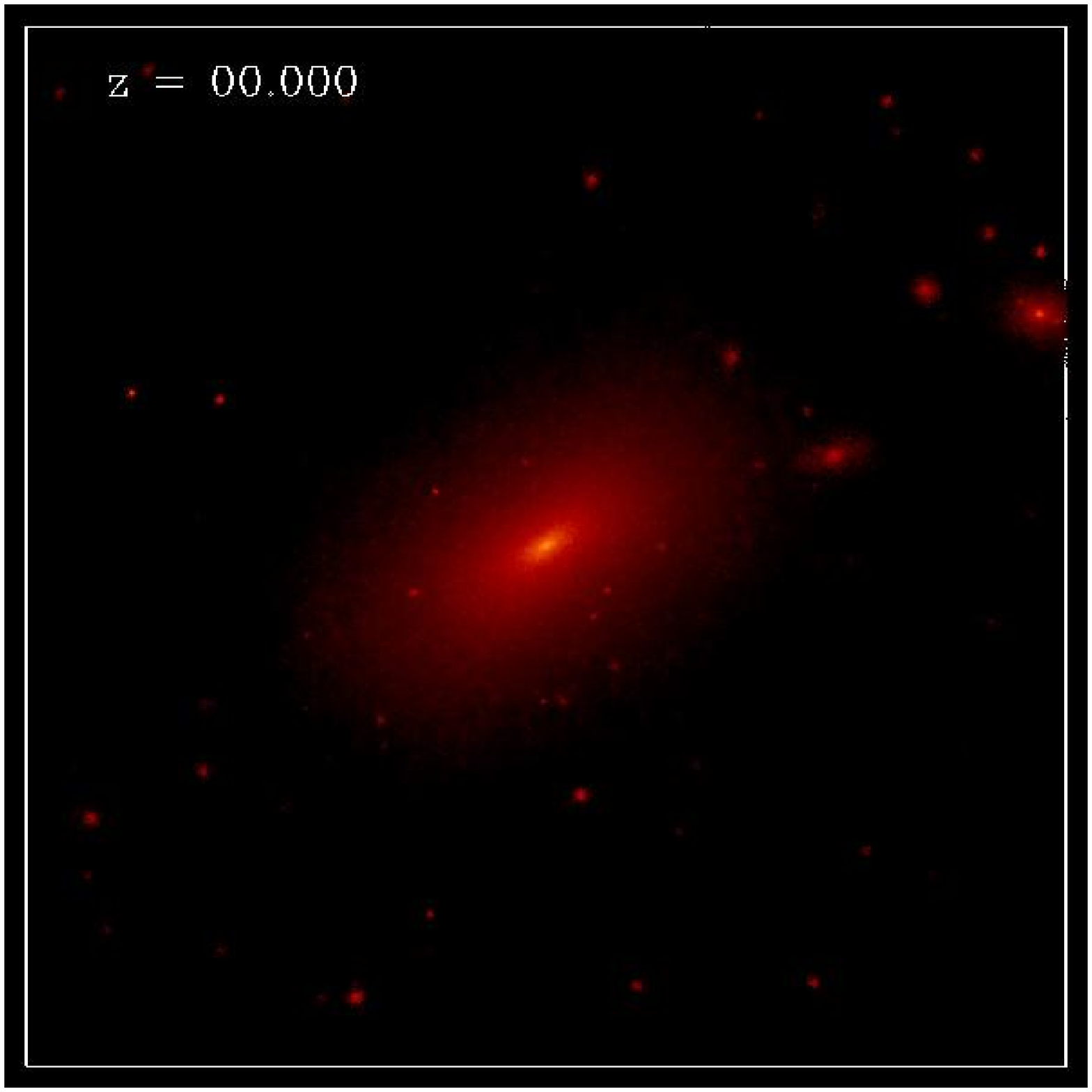}}
\resizebox{4.65cm}{!}{\includegraphics{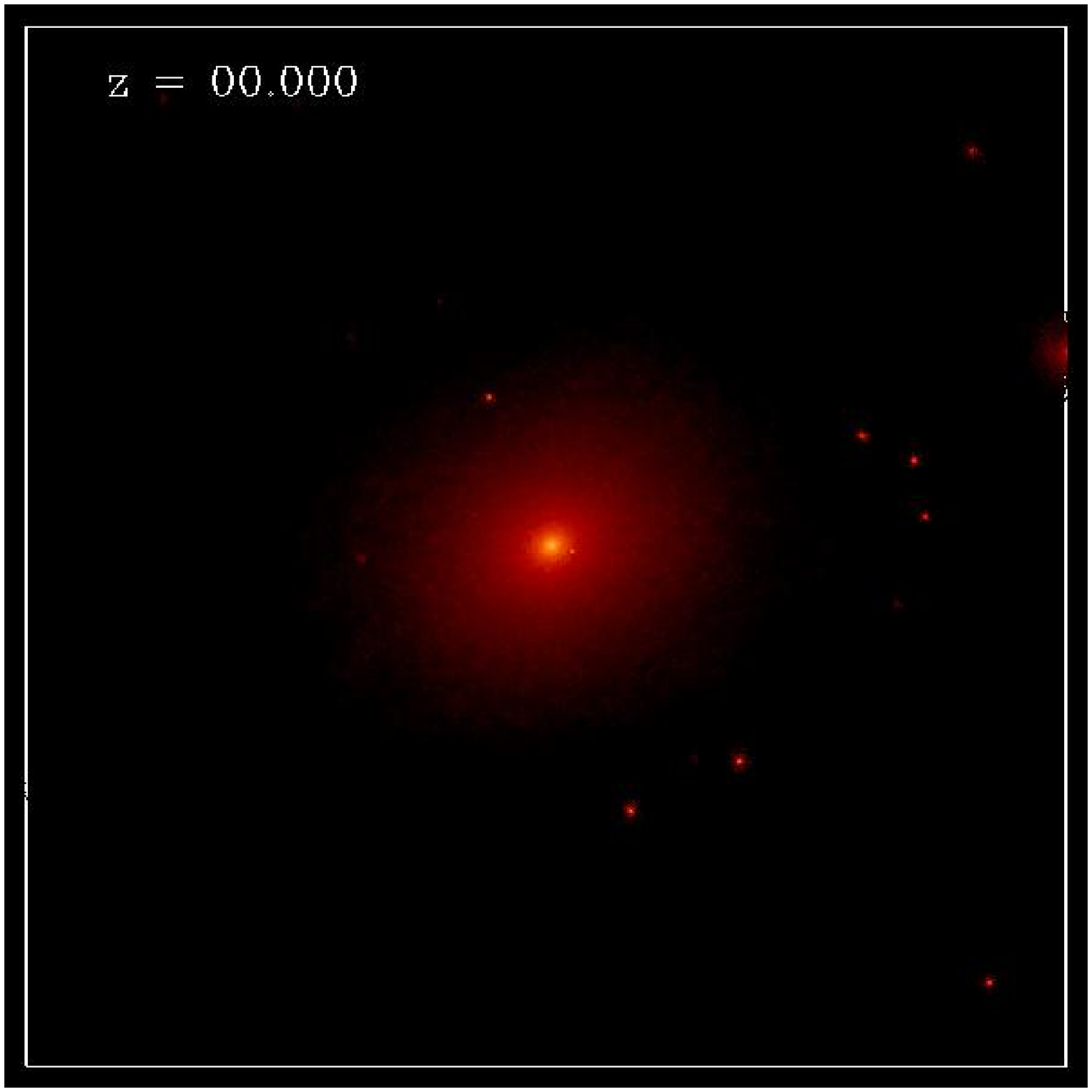}}
\end{center}
\caption{
\footnotesize
Example of DM halos at $z=0$ without (left frame) and with (right frame) baryons.
The frames are about 1~Mpc on the side. The halos have been evolved from identical
initial conditions (Romano-Diaz et al. 2009a).
}
\end{figure}

What is the source of the inner halos$'$ DM, say within $\tilde{R}_s$, and when do 
the inner halos form? Does the DM accumulation within ${\tilde R}_{\rm s}$ end with the
major mergers epoch (e.g., Wechsler et al. 2002). Comparison between the PDM
and BDM models reveals that only about $15\%$ and $33\%$ of the DM mass found within
this radius at $z=0$ is contributed by the subhalos, for PDM and BDM respectively.
The rest comes from the
`smooth' accretion. Here we, rather arbitrary, limit the `clumpy' (i.e., subhalo)
accretion to the clump-to-prime halo mass ratios in excess of $10^{-4}$ of 
the prime halo virial mass at $z=0$. This includes about $4\%$ and $20\%$ (PDM and
BDM) coming from the major mergers, i.e., the mass ratios of $\gtorder 1:3$, and 
$11\%$ and $13\%$ (PDM and BDM) from the minor mergers. Overall merger 
contribution to the
central mass falls slowly with the decreasing merger mass ratio. 
Second, while the DM density profile is established early (except in the central
few kpc), most of the DM is not confined to within the characteristic radii,
${\tilde R}_{\rm s}$ and $R_{\rm vmax}$ --- only $13\%$ and $20\%$ of DM found 
within these radii, respectively, are bound there. The majority makes much larger
radial excursions. Hence the buildup of the inner halos continues to $z=0$.
The flow of the unbound DM contributes about $80\%$ of the DM within 
${\tilde R}_{\rm s}$ after the merger epoch, but the net influx of this material 
is zero, and this region appears to be rather in a steady state. Most
of the DM particles within ${\tilde R}_{\rm s}$ reside there for a short
crossing time only. One hopes that these numbers are representative and reflect
the basics of the DM halo assembly.

Equally important for the disk evolution appears to be the highly anisotropic
flux of DM and baryons toward the center. The background potential of the DM
is strongly triaxial at the peak of the inflow.  
The shapes of DM halos respond to the presence of baryons by reducing their triaxiality
(e.g., Kazantzidis et al. 2004; Berentzen \& Shlosman 2006) and a number of factors
are involved (Fig.~3). Both, the clumpiness of baryons, their central concentration 
and out-of-phase response to the DM driving lead
to a decrease in the halo flatness and to a loss of its equatorial ellipticity, 
when the
disks form (e.g., Shlosman 2007). Moreover, the halo figure tumbling appears 
insignificant both with and without baryons, irrespective of whether it is located in
the field (i.e., isolated) or in a denser environment of a filament, based on 
$\sim 900$ snapshots (Heller et al. 2007; Romano-Diaz et al. 2009a), 
which agrees well with Bailin \& Steinmetz (2004) estimates.
The angular momentum acquired by halos
is channeled into the internal streaming rather than figure rotation. Such an 
exceedingly slow tumbling
of DM halos has important dynamical consequences for growing disks by moving 
the halo's outer inner Lindblad resonance to very large radii, beyond  
$R_{\rm vir}$.
 
\section{DM Halo Relaxation Beyond Virialization} 

While the DM halos are being virialized, this process is clearly
incomplete because a cosmological halo is an open system. In addition,
when the density profile is established early, oscillations of the central potential
decay fast, except during major mergers. This limits the
efficiency of the violent relaxation process (Lynden-Bell 1967). 
However, the ongoing accretion of a clumpy material can steer the halo,
especially its outer region where they are favored
because of the material influx along the large-scale
filaments.  It is informative, therefore, to look
at the fate of the incoming subhalos and compare the PDM and BDM models.
For this purpose we use the $R-v_{\rm R}$ phase diagrams which allow to
quantify the substructure contributions, where $R$ is
the spherical radius (Fig.~4). 

\begin{figure}[t!!!!!]
\begin{center}
\resizebox{6.6cm}{!}{\includegraphics{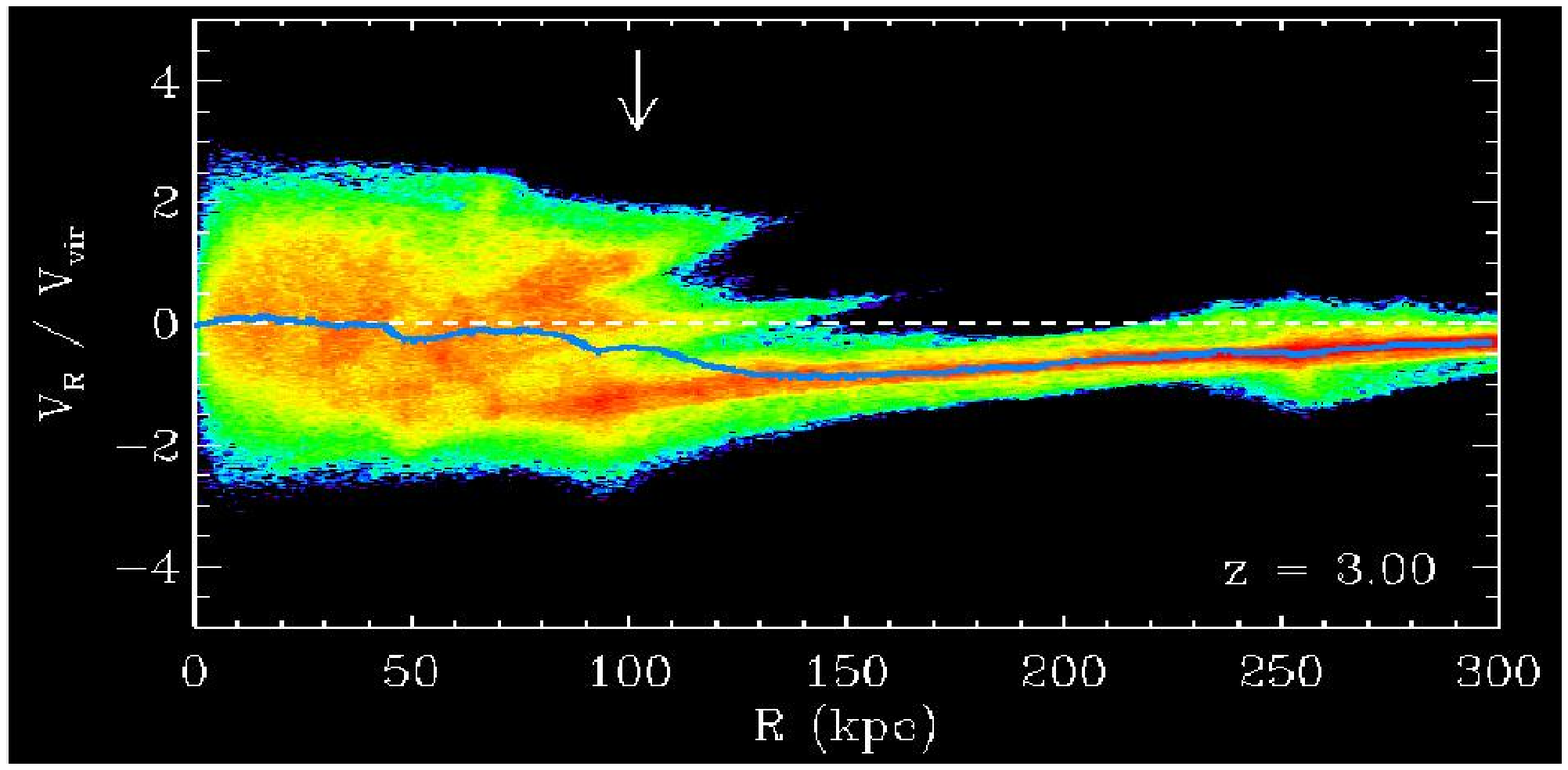}}
\resizebox{6.6cm}{!}{\includegraphics{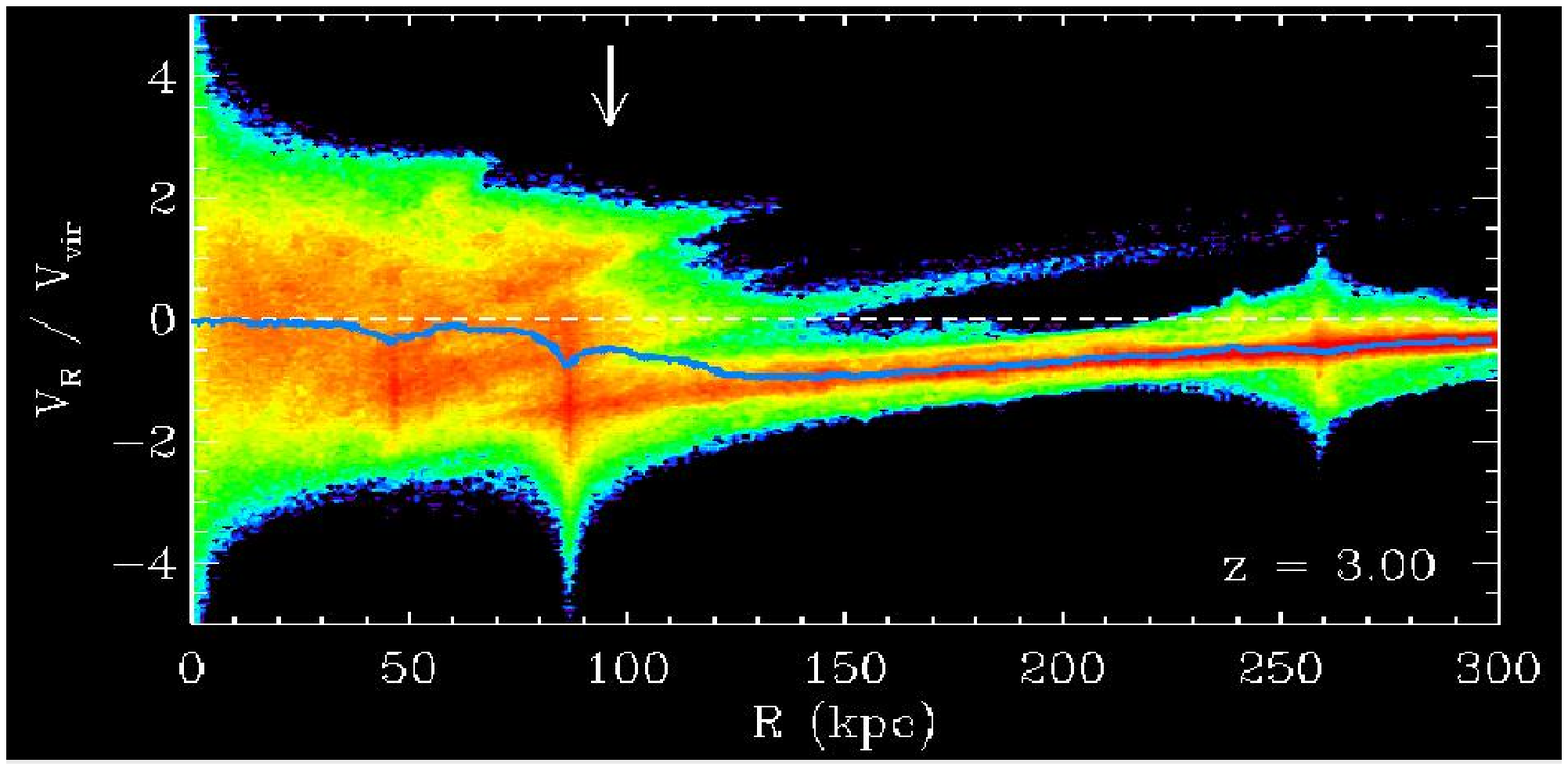}}
\end{center}
\caption{
\footnotesize
Evolution of the DM halo in the PDM (left) and BDM (right) models shown in the 
$R-v_{\rm R}$ plane at $z=3$, i.e., during the major merger epoch
and the appearance of ‘fingers.’ The colors correspond 
to the DM particle density on the $R-v_{\rm R}$ surface. The
vertical arrows show $R_{\rm vir}$, the dashed white line —-- $v_{\rm R} = 0$, 
and the blue line --- the average $v_{\rm R}$ at each $R$. Accretion along
the large-scale filament is seen as a stream at $v_{\rm R}<0$ 
(Romano-Diaz et al. 2009a).
}
\end{figure}

We define three types of subhalo relics, which partially represent various
stages of their dissolution. The (bound) subhalos can be easily seen because 
of their
vertical spikes (subhalo internal velocities) in Fig.~4. Those that are in the
process of tidal disruption are accompanied by inclined spikes (`fingers,'
partially bound).
Lastly, the tidally disrupted subhalos persist as streamers for a long time,
those represent $R-v_{\rm R}$ correlations (unbound). The ‘cold’ filament-driven
influx appears de-focused after passing the pericenter of its motion. As its 
constituents move out, their slowdown
in tandem with the tidal disruption lead to the formation
of shells that persist for a long time. Shells that form
later in time have larger outflow velocities and can cross
shells that formed earlier. For the survival of
these shells it is important that they form after subhalos
pass the pericenters of their orbits.

Both the spikes and fingers appear
longer by $\sim 2$ in the presence of baryons. The streamers survive for a
long time, and those that form after $z\sim 1$ survive till present. Hence,
the prime halo goes through inefficient relaxation beyond its virialization.
The ability of subhalos in the presence of baryons to penetrate deeper into
the prime halo and survive for a long time has also impications for the
disk evolution in the BDM models. The residual
currents in the halo continue to ‘mix’ the halo
environment even in the absence of large-amplitude variability
in the central potential.

\section{Disk-Substructure Interactions} 

\begin{figure}[t!!!!!]
\begin{center}
\resizebox{5.3cm}{!}{\includegraphics{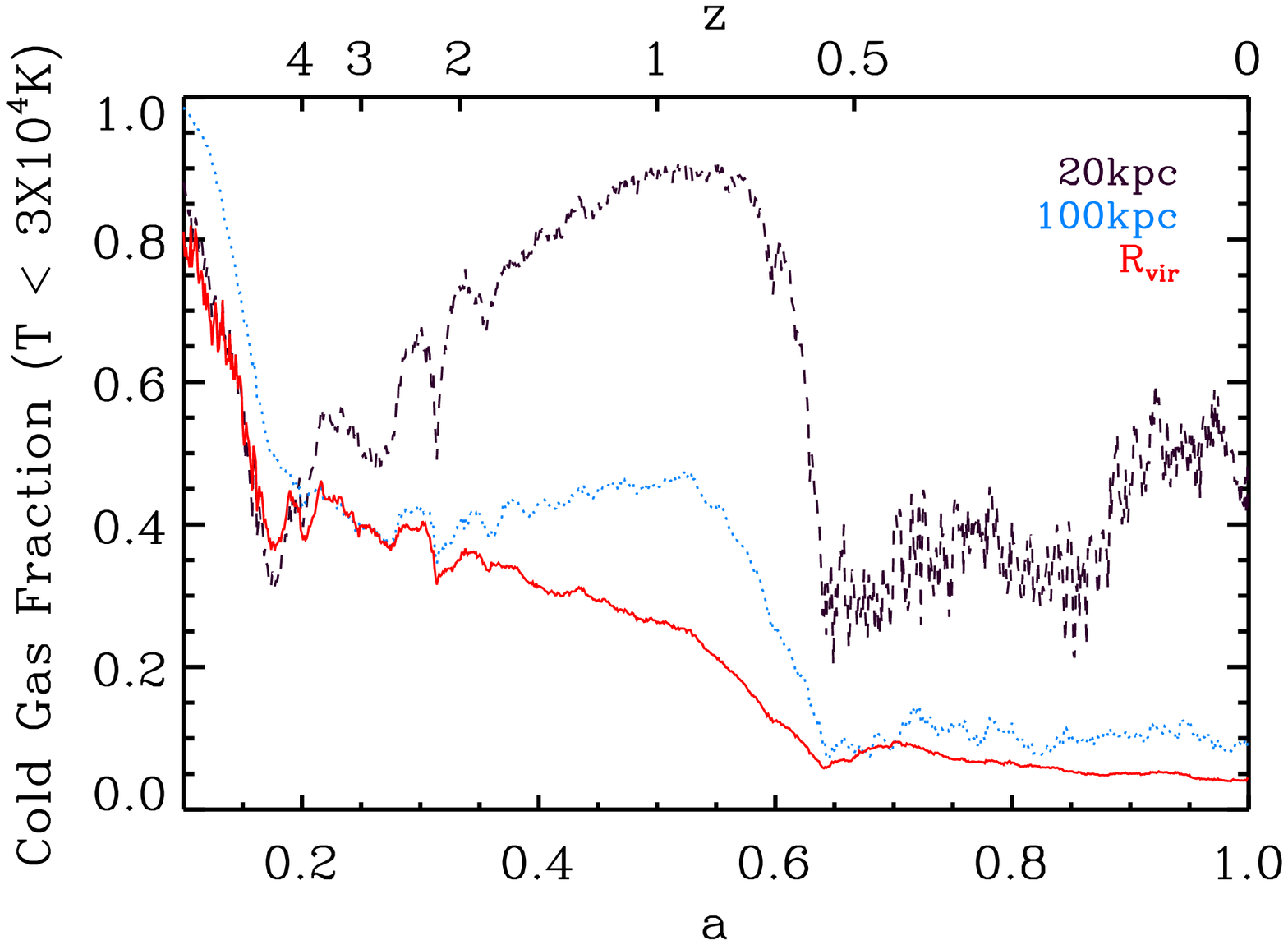}}
\resizebox{7.9cm}{!}{\includegraphics{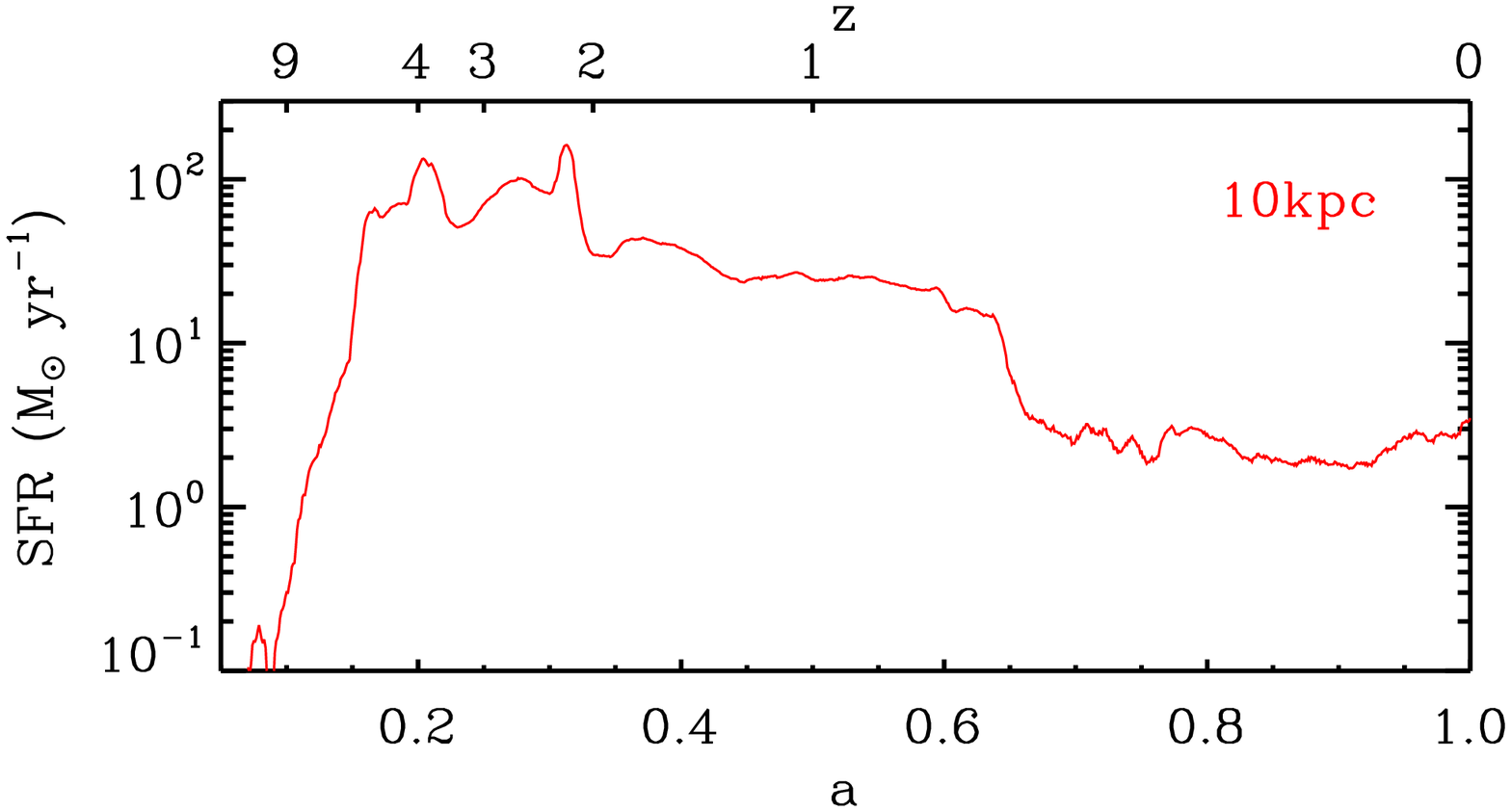}}
\end{center}
\caption{
\footnotesize
\underline{\it Left:\/} Fraction of the cold, $T\ltorder 3\times 10^4$~K
gas (of the total gas) within the inner 20~kpc (disk), 100~kpc and $R_{\rm vir}$ 
of assembling DM halo. The sharp decrease in the gas mass at $z\sim 0.5-0.6$ is 
related to the beginning of the subhalo influx along the large-scale filament 
and subsequent ablation from the disk; \underline{\it Right:\/} Star formation 
rates shown within the inner 10~kpc of the disk (Romano-Diaz et al. 2008a; 2009a).
}
\end{figure}

Evidence points to a variety of processes that affect
the disk assembly in a DM halo. Major mergers
act to destroy the forming disk, but if the system has a substantial
reservoir of cold gas and/or large-scale filaments continue to supply such
gas, the disk can be rebuilt at lower $z$. From general arguments,
the gravitational potential of the inner halo is not axisymmetric, and can include
the penetrating filaments, at least at high $z$. Under these conditions, 
the initially gas-dominated
disk will be strongly non-axisymmetric and turbulent. Bar formation is
favored and the properties of such gas/stellar bars differ from those
studied in the galactic dynamics. Because of the low central density concentration,
the first generation of bars will tumble with very low speeds (e.g., 
Heller et al. 2007; Romano-Diaz et al. 2008b). This property is crucial to
avoid the developing chaos (Shlosman 2009) which is expected to dissolve the 
bars formed in non-axisymmetric potentials because of the destructive interactions
between stellar and DM quadrupoles (El-Zant \& Shlosman 2002).  A rapid
growth of the disk brings along the bar growth as well, together with the
pattern speed-up. Moreover, the disk material, especially the gas but also the
stars, is aware of the large-scale orientation of the DM potential. One
should expect a nonlinear response of the disk shape to this perturbations.
Numerical simulations show that this leads to large distortions in the outer
disk, including prominent spiral arms, which are followed by a nearly
axisymmetric stage (Heller et al. 2007). Various waves are excited in the
disk and the beat phenomenon is frequently observed. Due to the presence of the 
gas, nested bar cascade is plausible and will lead to a rapid angular momentum
loss from the central regions, leading to the formation of a single massive
object (Begelman et al. 2006). 

The DM substructure can have a profound effect on the prevailing
morphology in the disk (Gauthier et al. 2006; Romano-Diaz et al. 2008b). Here, 
we mention two effects only: triggering
of galactic bars and cold gas ablation from the disk. The former
constitutes a finite perturbation and allows to circumvent  various
problems associated with the linear stage of the bar instability. Triggering
bars with galaxy perturbations has been known for some time (e.g., Byrd et al. 
1986; Noguchi 1987), but subhalos can make it even more efficient and, paradoxically,
less dependent on the environment. 

A potentially new effect is the ablation of the cold disk gas by subhalos, as
shown in Fig,~5a. It is clearly related to the penetrating encounters with
the subhalos. In addition to heating up the cold gas, the hot gas component is
driven out, in conjunction with the inner DM, as discussed 
in \S1. The immediate corollary of this process is a sharp, factor of 10
decrease in the star formation rate associated with the disk (Fig.~5b). This
can be a general stage in the evolution galaxies, when the star formation
is quenched by interactions with the DM substructure. 
 
\acknowledgements  
I thank the organizer, Shardha Jogee, for an interesting and stimulating
conference, and Emilio Romano-Diaz, Clayton Heller and Yehuda Hoffman for
collaboration on various topics discussed here. Support from NASA and NSF
grants is gratefully acknowledged.

\end{document}